\documentclass[12pt,twoside,a4paper,reqno]{amsart}
\usepackage{bulletinUPB_3}
\usepackage{graphicx}
\usepackage{amssymb}
\usepackage{eucal}
\usepackage{amsmath}
\usepackage[justification=centering]{caption}
\usepackage{float}
\usepackage{lipsum}
\usepackage{subcaption}
\usepackage{mathptmx} 
\usepackage{xcolor}
\info{A}{85}{1}{2022}
\title{A novel 3D sampling method of geological rock-core using X-ray fluorescence}

\author[1]{Alexandru ENCIU}

\address[1]{$^1$PhD Student, University Politehnica of Bucharest, Doctoral School of Applied Science, email: alexqi54@gmail.com}

\begin{document}
\pagestyle{headings}
\maketitle

\vspace{-10mm}
\begin{abstract}
{\it 
The current work describes a proof of concept of a 3D XRF scanner, which is able to perform elemental analysis over the cylindrical surface of geological rock-core and to reconstruct a 1D, 2D and a 3D elemental map of the scanned area. The presented method will reduce the time and cost for sample preparation, but also will reveal more information about the distribution of the elements over the surface.}
\end{abstract}

\begin{Keywords}
XRF 3D surface scan, 3D reconstruction, elemental analysis
\end{Keywords}




\section{Introduction}
    Since 1932 X-ray fluorescence (XRF) spectroscopy was considered a qualitative and quantitative elemental analysis by Hevesy, Coster and others who investigated in detail the method \cite{Handbook of X-Ray Spectrometry}. XRF analysis can cover a large range of elements (Z$>$10) with high resolving power, being able to detect concentrations down to a few ppm  \cite{Handbook of X-Ray Spectrometry,Handbook of Practical X-Ray Fluorescence Analysis,Principles and Practice of X-Ray Spectrometric Analysis}.\par 
For thin samples (thickness below 1 $\mu$m), there is linear dependency \cite{Handbook of X-Ray Spectrometry} between the intensity of the measured characteristic X-rays after the exposure to an X-ray generator and the number of atoms existing in the sample, i.e. the concentration of that species in the sample. On the other hand, the situation becomes more complex when a thick sample is analyzed due to  the interaction of the characteristic radiation with the sample matrix \cite{Handbook of X-Ray Spectrometry}.  The notion of matrix refers to all elements present in the sample except for the element which is analyzed. In standard practice samples are either homogeneous or special preparation is applied to minimize the anisotropic structures \cite{Handbook of X-Ray Spectrometry}. Classical protocols usually involve a standard material \cite{Handbook of X-Ray Spectrometry,Handbook of Practical X-Ray Fluorescence Analysis,Principles and Practice of X-Ray Spectrometric Analysis} with a similar matrix. In the absence  of a standard material or an extended analysis which determines the matrix coefficients, the measurement can only give qualitative information, which for some cases may suffice.  Mapping the relative concentrations of various elements  may offer a valuable  estimate for the sample topography that represents a fair hint within any geological analysis.\par  
    Another challenge that one should consider is represented by not having a well defined geometry. The scanned surface should be flat in order to maintain constant distances between detector, sample and X-ray generator and the roughness may also affect the results. The standard method in geology currently involves cutting the cylindrical rock-core sample in half exposing a flat rectangular surface which is suitable for 2D scanning \cite{ITRAX,M4,Bruker}. This approach requires additional cutting equipment, it is time consuming and invasive.\par 
The most used rock-core XRF scanner on the market is ITRAX\textregistered, made by company Avaatech, which is an multi-function core scanner due to its capability to perform $\mu$-XRF analysis and $\mu$-radiography on half-rock cores \cite{ITRAX}. The XRF analysis is performed by irradiate the sample with an X-ray fan-beam which is narrow, 100 $\mu$m, on the scanning direction and wider, 5 mm, on the perpendicular one in order to obtain high positioning  resolution over the geological layers and a good statistic in the same time. This type of scan leads only to 1D elemental maps, stratigraphic images. The 2D elemental maps are not possible to achieve with this kind of equipment. Usually if a 2D elemental map is required, a scanner as M4Tornado\textregistered \,made by Bruker is used \cite{M4,Bruker}. This type of scanner provides high resolution images due to the X-ray optics and microfocus X-ray tube. The major disadvantages are the limited size of the sample and the required flatness of it.\par
   The aim of this work is to show a proof of concept device which is able to scan the cylindrical shape of a rock core using XRF and to reconstruct an elemental distribution map over the sample. The next section describes the device and the working concept and the results in 1D, 2D and 3D representation are discussed in the third section.\par
    

\section{Materials and Methods}
    The proposed method consists of a 3D scan of the rock-core which allows for a 3D reconstruction of the superficial distribution of elements in the sample. This scan would serve as a preliminary map containing relative intensities of the elements and does not involve any preparation of the sample, reducing the costs and the complexity of the analysis. \par
    
    The sample used to commission the proposed method is a piece of breccia (Figure \ref{fig6A}) with a diameter of 60 mm and a height of 80 mm. Breccia are sedimentary rock formations which have fragments larger than 2 mm, and are formed where broken debris is accumulated (e.g base of the outcrop). The accumulated fragments are bound together by a mineral cement or by a matrix of smaller particles that fills the spaces between larger fragments \cite{Breccia}.\par
    
\begin{figure}[ht]
  \centering
  \includegraphics [width=\textwidth]{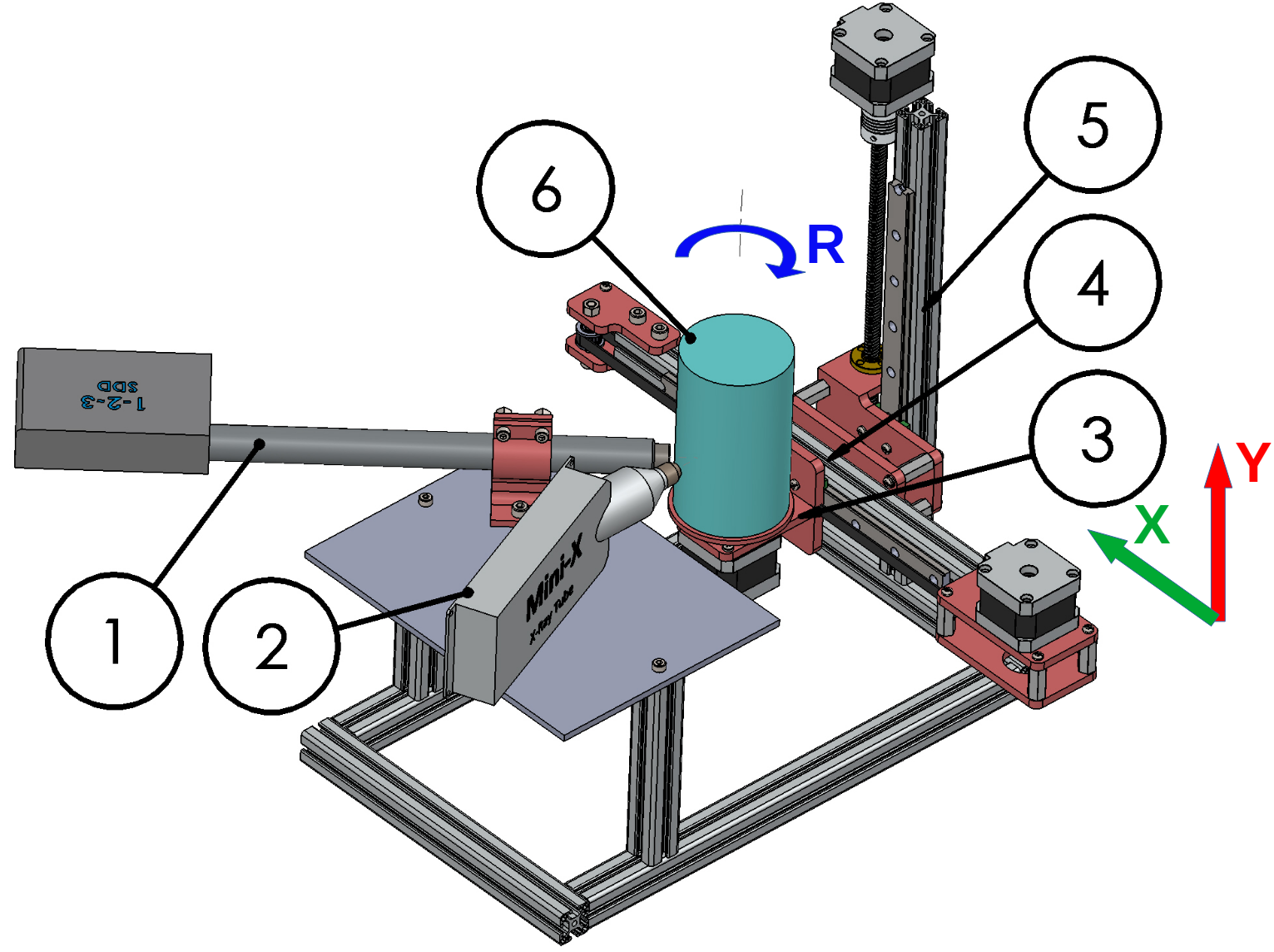}
  \caption[width=\textwidth]{CAD drawing of the proposed experimental setup.\par 1. X-Ray Detector; 2. X-Ray Generator; 3. Rotary table R; 4. Translation axis X; 5. Translation axis Y }
  \label{fig1}
\end{figure}

In order to perform a 3D scan over the rock-core surface, a custom experimental setup was built (Figure \ref{fig1}).The experimental setup consists of: an X-ray Detector, an X-ray generator and a mechanical sample manipulator which can perform translation movements of the sample over the X and Y axis and rotation over the R axis. The X-ray detector is an Amptek Si-PIN which has a resolution of 139-190 eV corresponding to the energy range of 1-30 keV and an optimal count rate of 30k cps. The X-ray generator used is an Amptek Mini-X with Ag target, 40 kV maximum acceleration voltage and 200 $\mu$A electron beam current. In order to obtain a high scanning resolution, the X-ray generator was collimated with a 0.5 mm copper aperture. The controlled movement of the sample is achieved by using a motorized axis system which was built with recycled components from old printers. The mechanical repeatability of the system is $\pm$75 $\mu$m on the X and Y axis and $\pm$0.2 deg on the R axis. The high positioning resolution was archived by using stepper motors and 32 bit micro-stepping drivers. The motors drivers are controlled with an Arduino UNO\textregistered \, \cite{Arduino} development board linked via USB to PC and the system can be controlled remotely.\par

    The software integration of the entire experimental set was performed using the visual programming language LabView\textregistered \, \cite{LabView}. The software allows separate control of each component and its parameters. The scanning sequence and data aquisition (DAQ) are performed automatically without the need for human intervention during the process.\par
	


   
Without the need of a preliminary preparation of the sample, the analysis starts by placing the sample on the rotating table and choosing the scanning geometry depending on the symmetry of the sample. In this case, the scan was done with cylindrical symmetry, rotating the sample uniformly while moving it on the vertical axis. The input parameters for DAQ are: X-ray generator voltage and current, X-ray detector integration time, the desired pixel size, the sample diameter and height. The angular step $\theta$ was calculated with Eq. \ref{eq1}.

\begin{equation}
\label{eq1}
\theta=2\arcsin(\frac{d}{2r})
\end{equation}
\\
where $d$ is the pixel size and $r$ is the radius of the sample. After each complete rotation, the sample is moved on the Y axis with distance $d$, in order to maintain a square pixel, and the process is repeated until the entire surface of the sample is scanned.\par
 The sample used for this work was scanned at a speed of 5 mm/s with a pixel size of 0.5 mm x 0.5 mm. The voltage and beam current of the X-ray generator are 35 kV and 150 $\mu$A. The distances between X-ray generator-sample and X-ray detector-sample are 10 mm and 8 mm with an angle of 45 deg between detector and generator.\par
   For the current work a custom analysis code was made in LabView\textregistered\, in order to process the large number of acquired spectra. The main steps for the data analysis are: data filtering for noise reduction, energy calibration, peak finding, detector efficiency correction, 1D, 2D and 3D elemental map reconstruction.\par
The acquired spectra are affected by statistical noise. In order to reduce it, a high-frequency cutoff filter $h(x)$ was convoluted with the pulse height spectrum $f(x)$. In the frequency domain, this type of filter is applied  as the product between Fourier transform of the pulse-height spectrum $F(u)$ and the Fourier transform of the digital filter $H(u)$  (Eq. \ref{eq2}).\par
    
\begin{equation*}
\begin{split}
F(u)=\frac{1}{n}\sum_{x=0}^{n-1}f(x)\exp{\frac{-j2\pi u x}{n}}\\
j=\sqrt{-1};\;u=0, 1...(n-1);\;x=0, 1...n;\\
H(u)= \begin{cases}
1,\;\;u\le u_{crit} \\
0,\;\;u>u_{crit}
\end{cases}
\end{split}
\end{equation*}

\begin{equation}
\label{eq2}
G(u)=F(u)H(u)
\end{equation}
\\
where $n$ is the number of channels, $u_{crit}$ is the cutoff threshold and $G(u)$ is the filtered spectrum in the Fourier space. The return to the normal space can be achieved by applying the inverse transform (Eq. \ref{eq3}).\par


\begin{equation}
\label{eq3}
g(x)=\sum_{u=0}^{n-1}G(u)\exp{\frac{j2\pi u x}{n}}
\end{equation}
\\
where $g(x)$ is the filtered pulse-height spectrum.\par
    The energy calibration was done by measuring spectra over the standard samples of Ca, Fe, Ni, Cu and Au with a purity of 99.99 \%. The calibration process involves identification of the spectral lines of each element, plotting the energy of the peak in function of the peak location channel and identifying the parameters a and b of the Eq. \ref{eq4} by making a linear fit over the plotted data.\par


\begin{equation}
\label{eq4}
E(x_{Channel})=ax_{Channel}+b
\end{equation}
\\
where $E(x_{Channel})$ is the energy which corresponds to a certain channel.\par
    The automatic peak finding method used in the analysis involves transformation of the spectrum in a way that the contribution of the continuum, background, part can be removed. The algorithm for the transformation performs the first and second derivative of the spectrum. The crossing of the X axis of the first derivative and the position on X axis of the minimum of the second derivative are giving a good approximation of the peak maximum position. With obtained information about peak position a spectrum fit can occur by using a Gauss distribution for peaks and a $3th$ grade polynomial for the background. This fit is required in order to determine the peak area ($I_{measured})$.\par
The measured intensity needs to be corrected for the energy dependent efficiency, $\epsilon(E)$, of the X-ray detector as in Eq. \ref{eq5}. The resulting corrected intensity, $I_{corrected}$, is closer to the mass fraction values of each element.\par


\begin{equation}
\label{eq5}
I_{corrected}=\frac{1}{\epsilon(E)}I_{measured}
\end{equation}

    The 1D representation is made by summing all spectra from each scanned layer and plot the value of the $I_{corrected}$ of a certain element in function of the layer height (Figure \ref{fig7}).\par
The final 2D elemental map is reconstructed with the $I_{corrected}$ values in each point $(\varphi,h)$. On the X and Y axis of the plot, the angle $\varphi$ and the height $h$ of the acquired point are represented with the value of corrected intensity ($I_{corrected}$) of desired spectral line on the Z axis (color scale) Figure \ref{fig4}.\par
    The 3D representation is made by plotting data in the cylindrical coordinate system $(r,\varphi,h)$ , where r is the radius of the rock-core. The 3D surface is made by building a mesh over the plotted points Figure \ref{fig6B}.\par



\section{Results and Discussion}

    In this section will be presented the results of the tabletop XRF scanner which scans the surface of the cylindrical rock-core in a manner that allows with collected data to reconstruct 1D, 2D and 3D elemental maps. 

\begin{figure}[H]
  \centering
  \includegraphics [width=\textwidth]{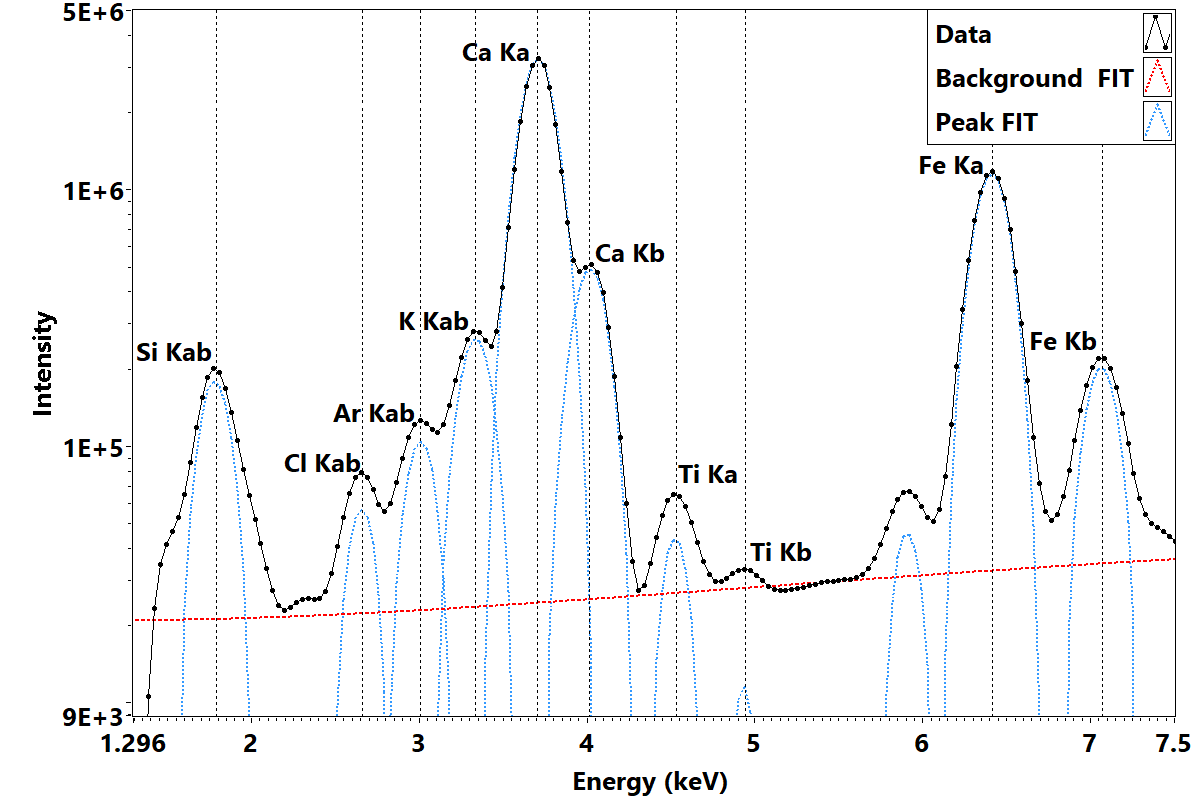}
  \caption[width=\textwidth]{Spectra summation of all measured points}
    \label{fig3}
\end{figure}

    The sum of all measured spectra is illustrated in logarithmic scale in Figure \ref{fig3}. The elements are identified and the peaks and background are fitted. This type of representation allows a better assembly view of the elements present in the sample.\par


\begin{figure}[ht]
  \centering
  \includegraphics [width=\textwidth]{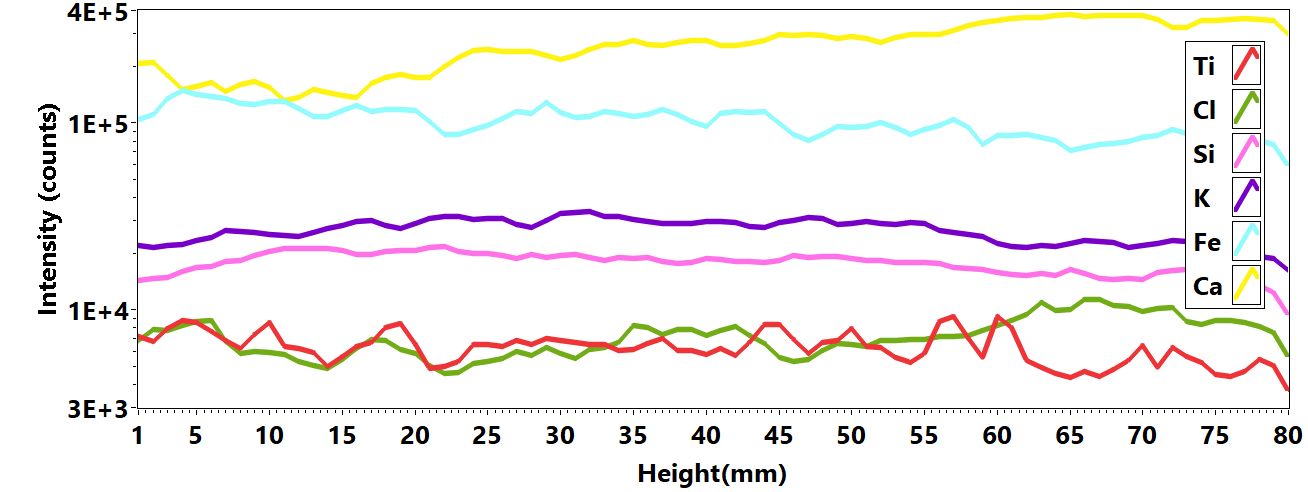}
  \caption[width=\textwidth]{Stratigraphic representation of the elements in the analyzed sample}
    \label{fig7}
\end{figure}

Figure \ref{fig7} displays the element profiles along the rock-core helping to understand the geochemical content of the sample in each geological layer. This type of representation is the most required in the rock-core analysis domain \cite{ITRAX}.\par


\begin{figure}[H]
  \centering
  \includegraphics [width=\textwidth]{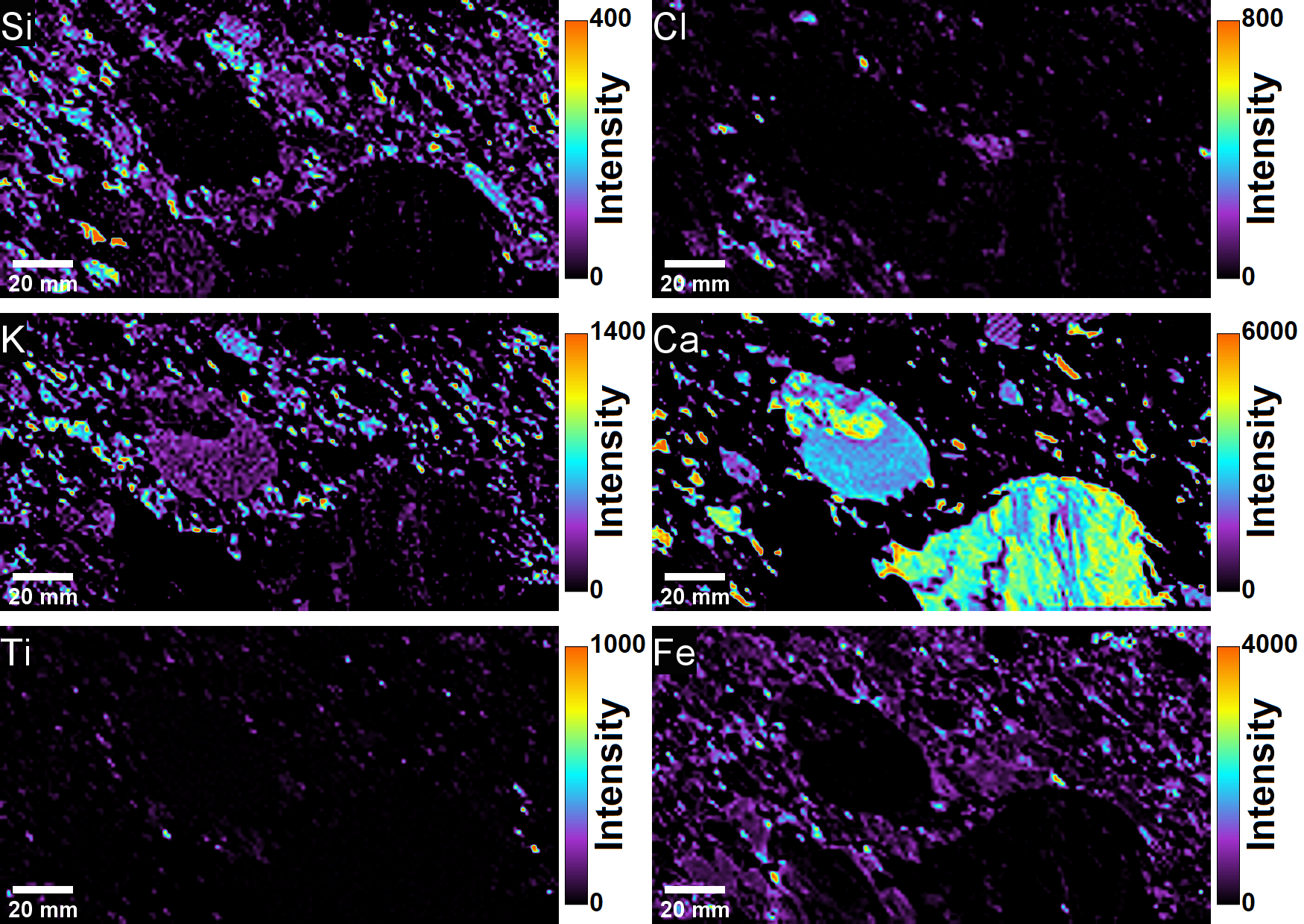}
  \caption[width=\textwidth]{2D elemental map reconstruction for Si, Cl, K, Ca, Ti and Fe}
    \label{fig4}
\end{figure}

In Figure \ref{fig4} the reconstruction of the 2D elemental maps for Si, Cl, K, Ca, Ti and Fe are made by using the K emission line of each element. The color of the pixel indicates how the abundance of the element is varying in the sample. Dark-purple means that the abundance is very low, while red means that the abundance is high. This type of representation allows measurements of the grain size of the sample, granulometry.

\begin{figure}
     \centering
     \begin{subfigure}[b]{0.3\textwidth}
         \centering
         \includegraphics[scale=0.07]{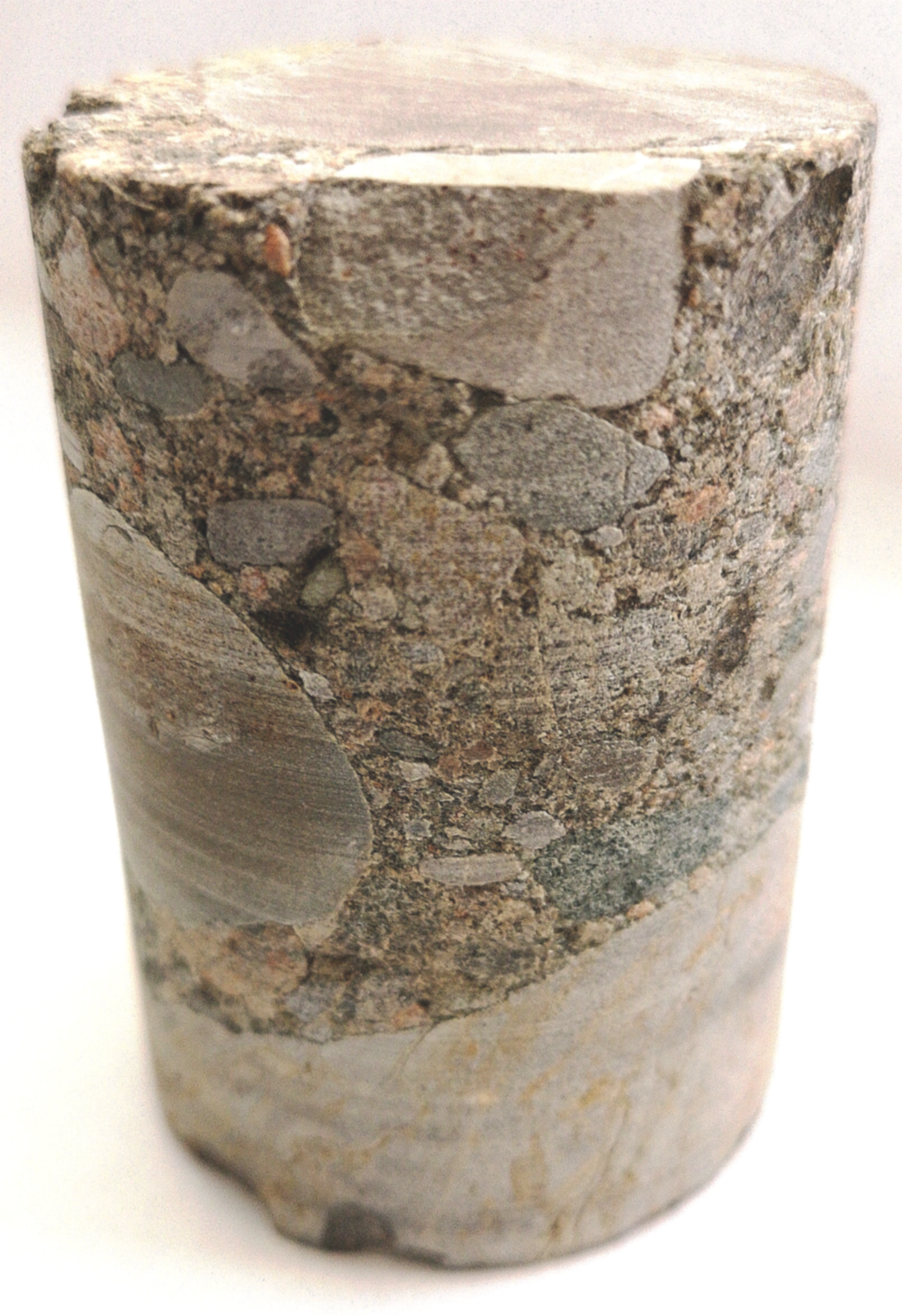}
         \caption{}
         \label{fig6A}
     \end{subfigure}
     \begin{subfigure}[b]{0.3\textwidth}
         \centering
         \includegraphics[scale=0.171]{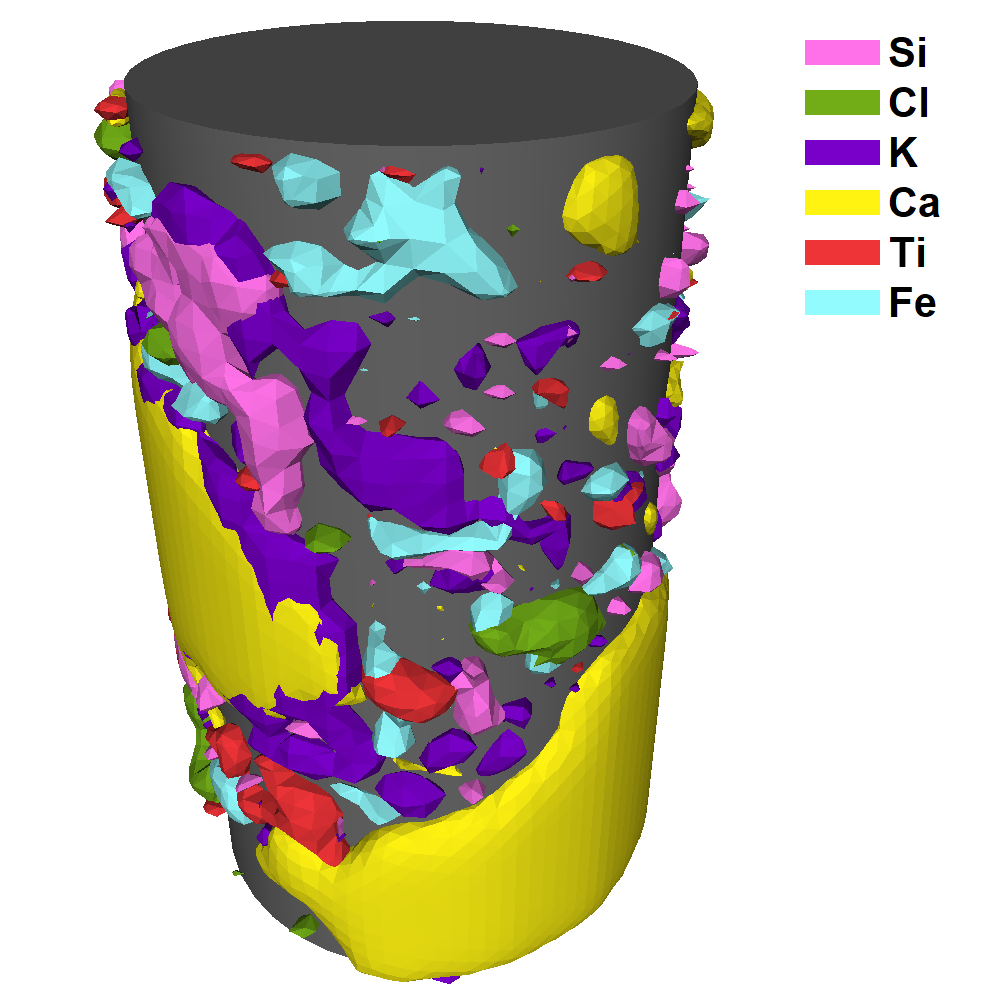}
         \caption{}
         \label{fig6B}
     \end{subfigure}
     \label{fig6}
     \caption{A. Breccia rock-core sample with a diameter of 60 mm and height of 80 mm B. 3D reconstruction of Si, Cl, K, Ca, Ti and Fe over rock-core surface }
\end{figure}

The 3D reconstruction (Figure \ref{fig6B}) was made for all elements of the sample in order to obtain a more acquired view of the entire surface. For view purposes only, the mineral cement is not displayed in order to better view the elemental clusters over the surface which leads eventuality to the granulometric measurements.\par


    The difficulties of the presented method consist in centering the sample on the rotary table and miss positioning on the Y axis due to the sample weight. The current design is not suitable for long samples, ideally a laser profiling system and a horizontal design needs to be made.\par




\pagebreak 
\section{Conclusions}
    The proposed method offers high quality images of rock-core surface at high scanning speed with minimal sample preparation and due to the small pixel size, the quality of the images was good and allowed to establish if any ROI needs furthermore analysis.\par
    
    The current method combines the scanning capabilities of the two devices (ITRAX\textregistered\, and M4Tornado\textregistered) in only one data collecting process by expanding the scanning sequence from 2D to 3D. In this way the collected data can be presented in various ways 1D, 2D and 3D in order to satisfy the needs of the user.\par    
By scanning the surface of the cylinder, the scanned area is 6 times larger than in the classical method, leading to greater statistics when 1D representation is performed. Another advantage of the surface scan is that it allows the user to view and measure the dimensions of the mineral clusters when 2D and 3D representations are used.\par



\end{document}